\documentclass[12pt]{article}
\usepackage{graphicx}
\begin{document}
\centerline{\Large \bf Percolation in high dimensions is not understood}

\bigskip
S. Fortunato$^1$,  D. Stauffer$^2$ and A. Coniglio$^3$

\bigskip
$^1$ Faculty of Physics, Bielefeld University, D-33615 Bielefeld, Germany

fortunat@physik.uni-bielefeld.de
 
\bigskip
$^2$ Institute for Theoretical Physics, Cologne University, 

D-50923 K\"oln, Euroland

stauffer@thp.uni-koeln.de

\bigskip
$^3$ Dipartimento di Fisica, Universit\`a di Napoli `Federico II', 

Via Cintia, I-80126 Naples, Euroland

Antonio.Coniglio@na.infn.it
\bigskip

Abstract: The number of spanning clusters in four to nine dimensions does not 
fully follow the expected size dependence for random percolation. 
\bigskip

Researchers were interested already long ago in percolation theory above the 
upper critical dimension of six \cite{aharony,coniglio}, and we followed
\cite{stauffer}.  At the percolation threshold \cite{grass}, there is a
theoretical consensus that the number $N$ of spanning clusters stays
finite with increasing lattice size below $d=6$ dimensions, and
increases with some power of the lattize size above six dimensions,
for hypercubic lattices of $L^d$ sites \cite{lda}. Andronico et
al.\cite{fortunato}, however, have worrying data in five dimensions
showing an increase of $N$ with increasing $L$. Thus we now check this
question.

One Fortran program, available from stauffer@thp.uni-koeln.de, checks if a
cluster spans from top to bottom and uses free boundary conditions in this and
one other direction, while helical boundary conditions are used in the 
remaining $d-2$ directions. The spanning properties are known to depend on 
boundary conditions and thus no quantitative agreement with \cite{fortunato}
is expected. In three dimensions the average $N$ is about 0.4 for $L = 7$ to 
101, roughly independent of $L$ as predicted; that means there is often no 
spanning cluster. Figure 1, however, shows for $d = 5$ an increase of 
$N$ with increasing $L = 3$ to 101. Figure
2 shows for $d=7, 8$ and $9$ an increase of the multiplicity
as $L^{1.65}$, $L^{2.49}$ and $L^{3.39}$, respectively. The points in Figs. 1 and
2 are averages over mostly 1000 runs.

The other Fortran program uses free boundary conditions in all directions 
and it is available from fortunat@Physik.Uni-Bielefeld.DE. Its results in Figs.3 and 4,
which refer mostly to a number of iterations
between 10000 and 50000, are 
qualitatively similar to Figs.1 and 2. However, one derives instead an increase 
of the spanning cluster multiplicity as $L^{0.97}$, $L^{1.53}$ and 
$L^{2.1}$ for $d=7, 8$ and $9$, 
respectively. We remark that this series of slopes is 
quite well reproduced by the simple formula $(d-5)/2$, which is 
not predicted by any theory and which, if true, would hint the existence
of infinite spanning clusters at threshold already in five dimensions.
In fact, even the trend of the 6D data is quite well reproduced by 
a power law with exponent 0.51, which is amazingly close to the 1/2 that one
would derive from the above mentioned formula. The 6D data points 
derived by the first program (Fig.1) can be instead better described by
a logarithmic law, in accord with theory: one sees an increase as $\log^2(L/2)$. 
The best fit exponents derived by the two
sets of data for $d=6$ to $9$ are listed in Table 1.

\vskip0.7cm

\begin{table}[h]
\begin{center}
\begin{tabular}{|c|c|c|}
\hline$\vphantom{\displaystyle\frac{1}{1}}$
& P.B.C. & F.B.C.   \\
\hline$\vphantom{\displaystyle\frac{1}{1}}$
6D & 0 ($\log^2(L/2)?$) & 0.51\\
\hline$\vphantom{\displaystyle\frac{1}{1}}$
7D & 1.65 & 0.97\\
\hline$\vphantom{\displaystyle\frac{1}{1}}$
8D & 2.49 & 1.53\\
\hline$\vphantom{\displaystyle\frac{1}{1}}$
9D & 3.39 & 2.10\\
\hline
\end{tabular}
\vskip0.3cm
\caption{Best fit scaling exponents of the spanning cluster multiplicity
with the lattice size L, corresponding to the mixed boundary conditions (P.B.C.)
of the first program and to the free boundaries (F.B.C.) of the second program.
The latter series is well described by the formula $(d-5)/2$.}
\end{center}
\end{table}

For $d=5$ both data sets show an analogous behaviour: the expected plateau is not reached even
at the largest lattice used, $L = 101$ for periodic boundary conditions, 
$L =70$ for free boundaries. Instead, the trend is quite well described 
in both cases by a logarithmic law.

As far as the comparison with theory is concerned, 
neither of the sets of exponents of Table 1
agrees with existing predictions.
Moreover, they do not agree either with the following plausible argument:
 
Let us assume that above the upper critical dimensionality the linear
dimension of the system $L$ does not scale asymptotically with the
correlation length $\xi$, instead it scales with a "thermodynamical"
length $\xi_T$. This length diverges as the critical point is
approached with an exponent $\nu_T = 3/d$ for percolation and $\nu_T =
2/d$ for Ising \cite{stauffer,luijten} models.

What is the meaning of this length $\xi_T$ ?  We believe
\cite{coniglio,lda,fortunato} that the number of incipient infinite
clusters $N_1$ in a region of linear dimension $\xi$ scales as
$$N_1 \propto \xi ^{d-6} \quad (d > 6) \quad.$$
The average distance $\xi_1$ between the "centers" of these clusters
is given by
$(\xi/\xi_1)^d \propto \xi ^{d-6}$.
Consequently
$\xi_1 \propto \xi^{6/d} \propto \xi_T$.
So $\xi_T$ is the average distance between the "centers" of the
spanning clusters in a region of linear dimension $\xi$.

How many spanning clusters are there in a region of linear
dimension $\xi_T$?  If the clusters did not interpenetrate one would
find only one cluster. However, since the clusters do interpenetrate
there are many more, depending strongly on the boundary conditions.
As first approximation we can assume that there are
$N_1 \propto \xi ^{d-6}$ spanning clusters.

Using the relation
$\xi_T \propto \xi^{6/d}$,  we obtain
$N_1 \propto \xi_T^{d(d-6)/6}$.
Since $\xi_T$ scales as $L$, we get the result
that the number of spanning clusters $N_1$ scales as
$$N_1 \propto L^{d(d-6)/6}$$
which gives the exponents 1.17 $(d=7)$, 2.67 $(d=8)$ and 4.5 $(d=9)$.
From Table 1 we see that if, on the one hand, the predictions for $d=7,8$ can 
be taken as possible interpolations of the two numerical values
we found, the results in nine dimensions (3.39, 2.10) seem to rule out this possibility,
being both sensibly smaller than the predicted value (4.5).  

Of course, one can always say that the simulated lattice sizes were too small,
but nevertheless the discrepancies are worrying.

\medskip
{\bf Acknowledgements:} This paper was partly written up while DS was
at Ecole de Physique et Chimie Industrielles, Lab. PMMH, in Paris; he thanks 
D. Tiggemann for help.  AC would like to acknowledge partial support from 
MIUR-PRIN 2002 and MIUR-FIRB 2002. SF acknowledges the financial support of
the TMR network ERBFMRX-CT-970122 and the DFG Forschergruppe FOR
339/1-2.

\newpage
  
\begin{figure}[t]
\begin{center}
\includegraphics[angle=-90,scale=0.42]{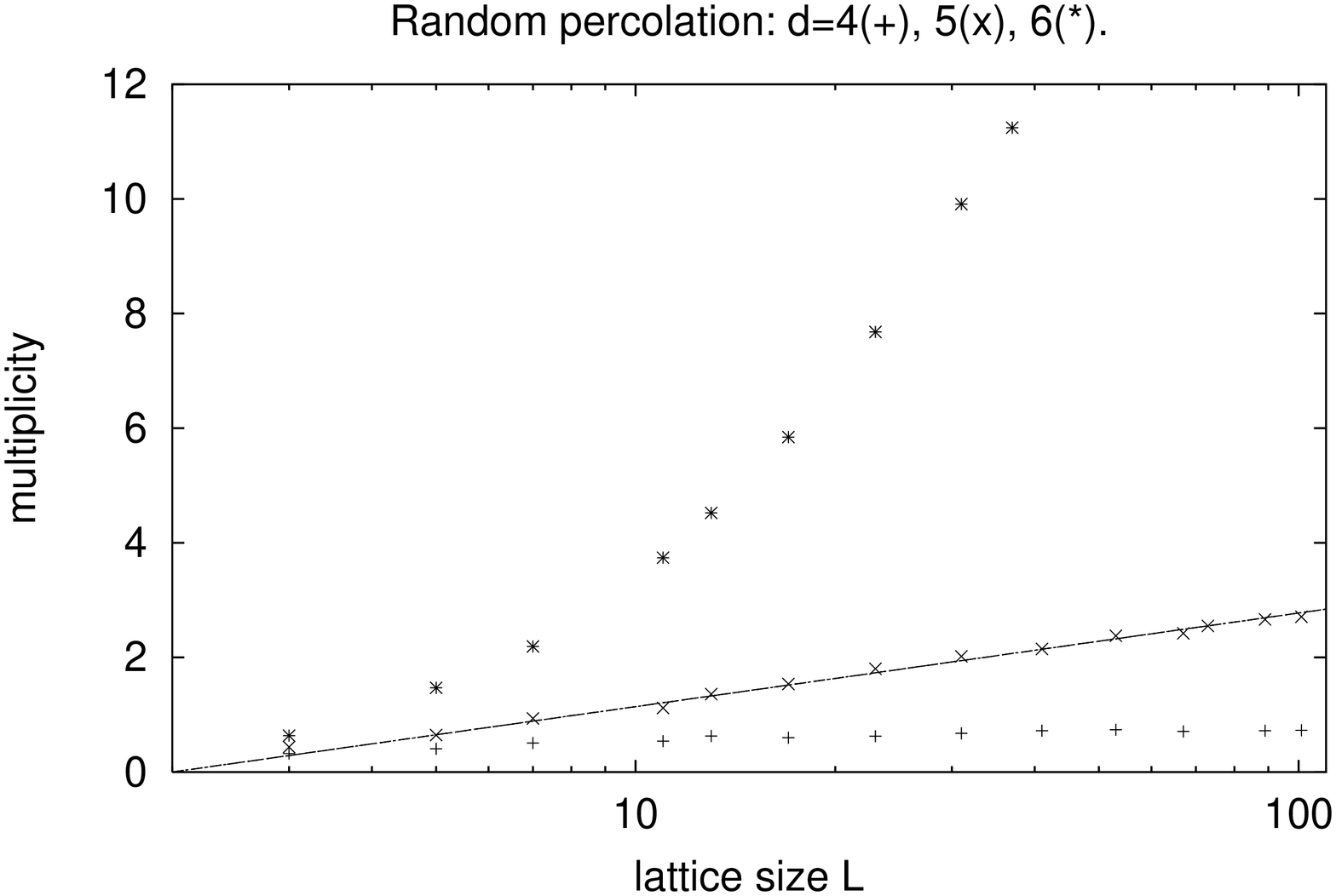}
\end{center}
\caption{Average number $N$ of spanning cluster versus linear lattice 
dimension $L$ in four, five and six dimensions. (Horizontal axis is
logarithmic.)
}
\end{figure}

\begin{figure}[b]
\begin{center}
\includegraphics[angle=-90,scale=0.42]{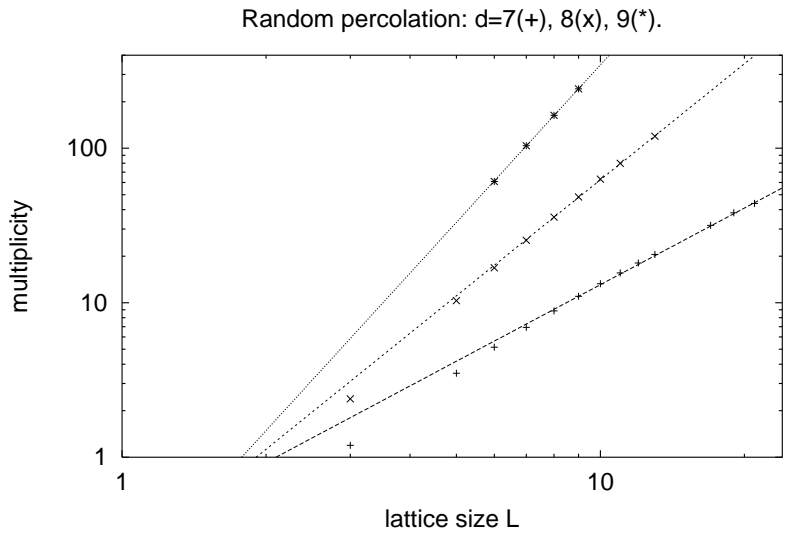}
\end{center}
\caption{As Fig. 1 but in seven, eight and nine dimensions. (Both axes are 
logarithmic.) The slight curvature suggests lower asymptotic slopes.
}
\end{figure}

\begin{figure}[hbt]
\begin{center}
\includegraphics[angle=-90,scale=0.42]{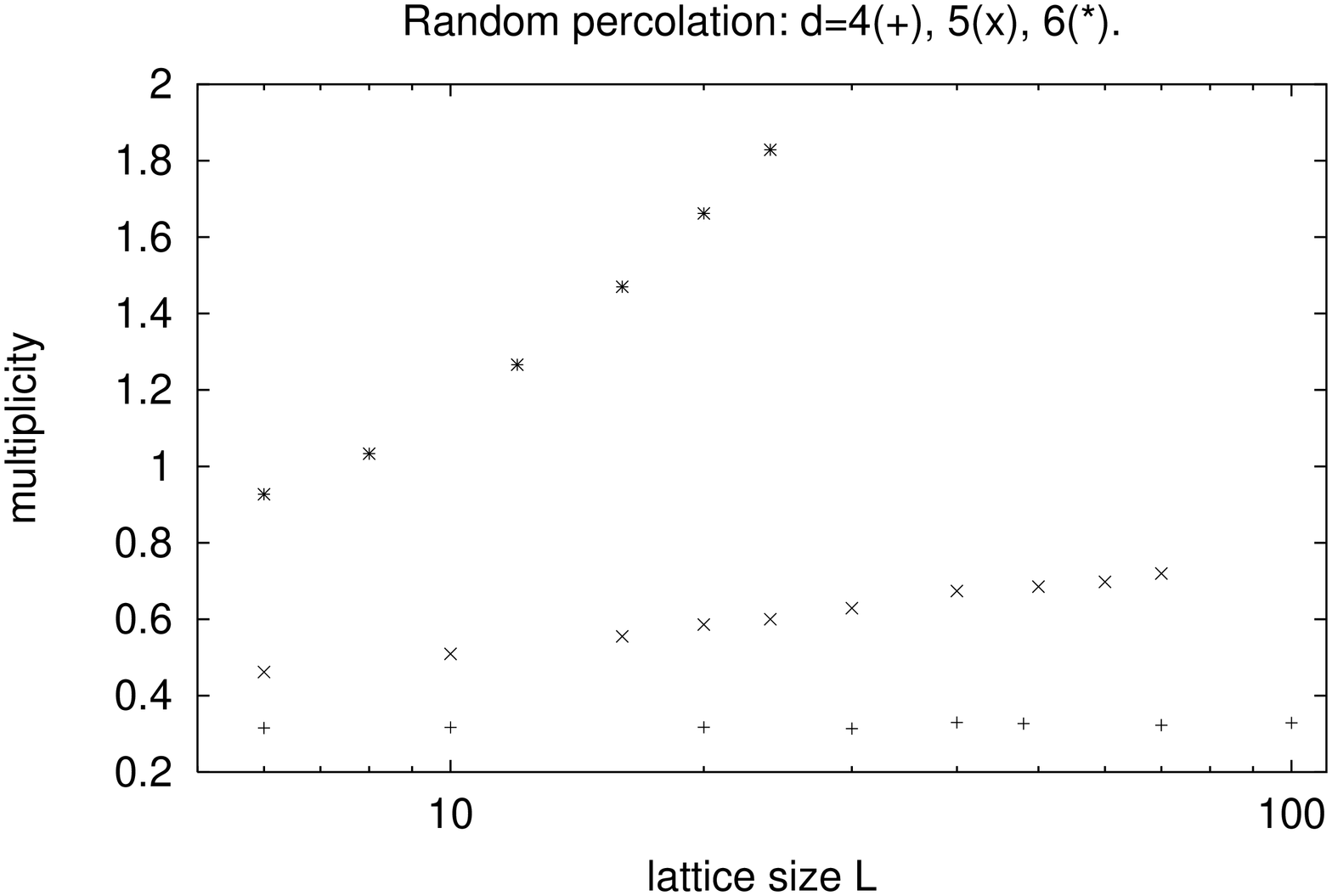}
\end{center}
\caption{As Fig. 1 but with free boundary conditions in all $d$
directions. }
\end{figure}

\begin{figure}[hbt]
\begin{center}
\includegraphics[angle=-90,scale=0.42]{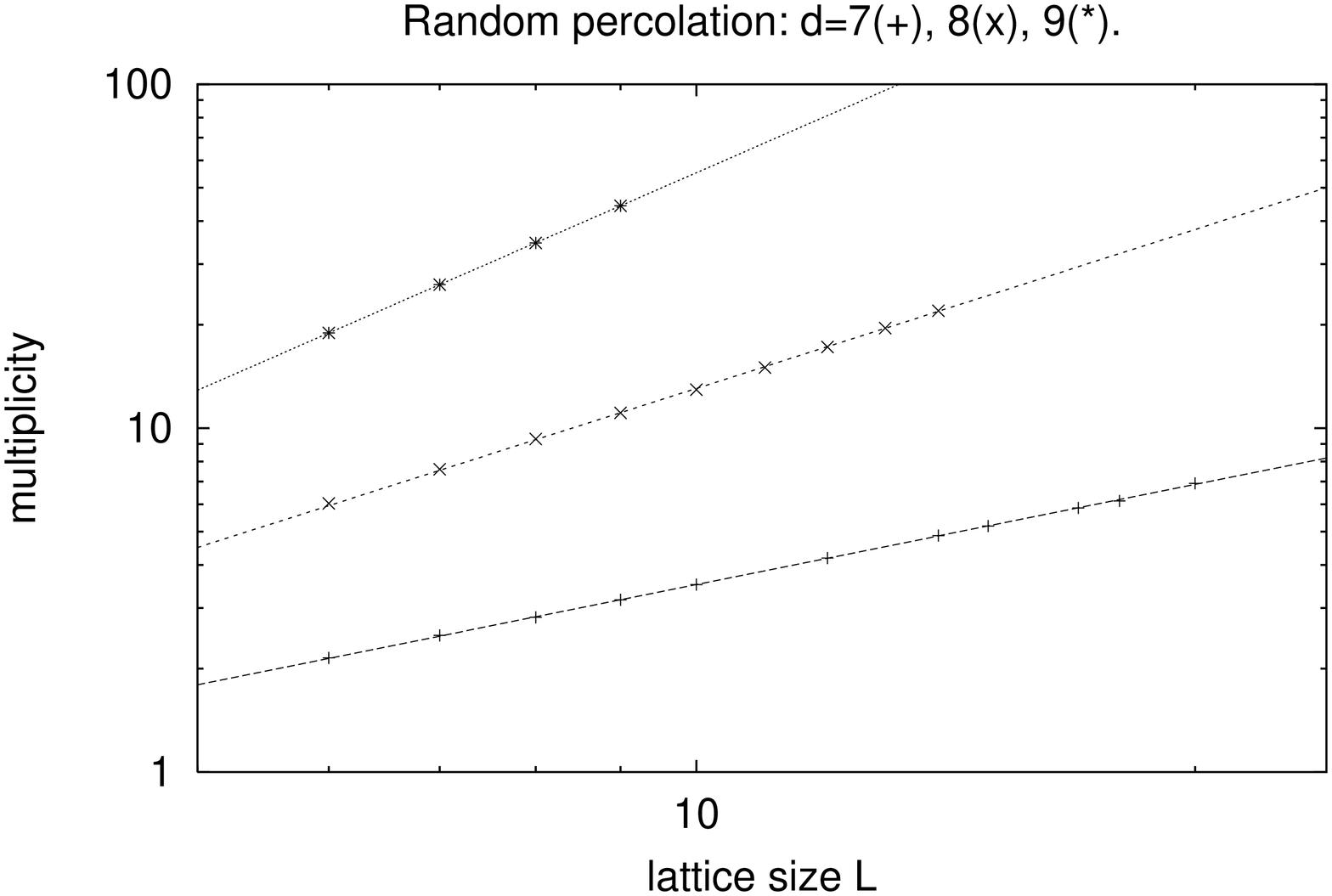}
\end{center}
\caption{As Fig. 2 but with free boundary conditions in all $d$
directions.}
\end{figure}

\end{document}